\documentclass[iop]{emulateapj}
\shorttitle{HD106906}
\shortauthors{Moore et al.}
\usepackage{amsmath,graphicx,natbib}
\usepackage{color}
\begin{document}

\title{Formation History of HD106906 and the Vertical Warping of Debris Disks by an External Inclined Companion}
\author{Nathaniel W. H. Moore\altaffilmark{1}, Gongjie Li\altaffilmark{1}, Lee Hassenzahl\altaffilmark{1}, Erika R. Nesvold, Smadar Naoz\altaffilmark{2,3}, Fred C. Adams\altaffilmark{4}}
\affil{$^1$ Center for Relativistic Astrophysics, School of Physics, Georgia Institute of Technology, Atlanta, GA 30332, USA}
\affil{$^2$ Department of Physics and Astronomy, University of California, Los Angeles, CA 90095, USA}
\affil{$^3$ Mani L. Bhaumik Institute for Theoretical Physics, University
of California, Log Angeles, CA 90095, USA}
\affil{$^4$ Physics Department, University of Michigan, Ann Arbor, MI 48109, USA}

\email{natemo13@gatech.edu}

\begin{abstract}
HD106906 is a planetary system that hosts a wide-orbit companion, as well as an eccentric and flat debris disk, which hold important constraints on its formation and subsequent evolution. The recent observations of the companion constrain its orbit to be eccentric and inclined relative to the plane of the debris disk. Here, we show that, in the presence of the inclined companion, the debris disk quickly ($\lesssim5$ Myr) becomes warped and puffy. This suggests that the current configuration of the system is relatively recent. We explore the possibility that a recent close encounter with a free floating planet could produce a companion with orbital parameters that agree with observations of HD106906b. We find that this scenario is able to recreate the structure of the debris disk while producing a companion in agreement with observation.
\end{abstract}

\section{Introduction} \label{sec:intro}

Classical theories of planetary system formation include the generally accepted core accretion model in which rocky cores form through planetesimal collisions followed by rapid accretion of gas from the solar nebula \citep{pollack_formation_1996}. Another hypothesis includes the gravitational instability mechanism, in which the solar nebula fragments through its own self-gravity into globs of gas and dust which then shrink and collapse to form giant planets \citep{boss_giant_1997}. However, as more exoplanetary systems are discovered, we are finding more unusual systems which seem to challenge our classical theories of planetary formation. For example, it's not quite clear what role external gravitational perturbers may play in the formation of planets and the subsequent architecture of the system they inhabit. Close flyby encounters with stars and brown dwarfs have been invoked to explain observed debris disk asymmetries such as the examples seen in $\beta$ Pictoris \citep[e.g.,][]{kalas_asymmetries_1995, ballering_comprehensive_2016}, HD141569 \citep{ardila_dynamical_2005, reche_observability_2008}, HD15115 \citep{kalas_discovery_2007} and HD100453 \citep{wagner_orbit_2018}. Close flyby encounters have also been used to explain unusual orbital architectures in planetary systems since flybys can perturb the orbits of objects in far distant orbits and leave unique features in their configuration \citep{cai_survivability_2019,batygin_dynamics_2020,li_fly-by_2019,li_flyby_2020,wang_planetary_2020}.

Debris disks are gas-poor disks of dust which orbit around their host star and can hold important clues in understanding the evolution of planetary systems. The detection of a debris disk indicates the successful formation of at least $>1$ meter objects during the protoplanetary disk phase since the dust in debris disks are continually replenished through collisions between these planetesimals \citep{wyatt_evolution_2008,matthews_observations_2014,hughes_debris_2018}. The architecture of an underlying planetary system can imprint its presence on disk material leaving distinct features in the disk's morphology \citep{wyatt_how_1999,lee_primer_2016,nesvold_circumstellar_2016}.

One unusual system, HD106906, contains an asymmetric debris disk in addition to a massive $11\pm2\mathrm{M_{Jup}}$ companion (HD106906b) at a large $732\pm30\mathrm{AU}$ projected separation \citep{bailey_hd_2014}. This companion is oriented $\sim21^\circ$ from the position angle of the disk mid-plane, which we view nearly edge-on \citep{kalas_direct_2015}. This orientation suggests that the orbit of the companion and asymmetric debris disk are not co-planar \citep{wu_magellan_2016}. Such an unusual system architecture makes HD106906 a rich opportunity for study. The existence of such a massive companion, far from its host binary and inclined relative to its debris disk, challenges the aforementioned classical theories of planetary formation due to the typically limited amount of material in protoplanetary disks at such large distances \citep{lieman-sifry_debris_2016,maury_characterizing_2019}.

HD106906 belongs to the 15 Myr-old Lower Centaurs Crux (LCC) subgroup within the Scorpius-Centaurus (Sco-Cen) OB association \citep{de_zeeuw_hipparcos_1999}. At the heart of the system lies a near equal mass binary with a short orbital period ($<100$ days) \citep{lagrange_narrow_2016}. More recent observations were able to detect orbital motion of the companion and further constrain its orbit \citep{nguyen_first_2021}. These recent constraints on the orbit of the companion confirmed the eccentric nature of the companion's orbit (eccentricity of $0.44^{+0.28}_{-0.31}$) as well as the misalignment of the companion's orbit and the asymmetric debris disk (mutual inclination of $36^{+27\circ}_{-14}$ or $44^{+27\circ}_{-14}$ depending on the orientation of the companion's orbit). Additional observation has constrained the obliquity of the companion and strongly suggest that the spin axis and orbit normal of the companion are misaligned \citep{bryan_obliquity_2021}. 

Recent observational studies have also revealed particular features of the asymmetric debris disk surrounding HD106906. Specifically, \cite{kalas_direct_2015} found that the disk extends roughly to $\gtrsim500$ au. Following earlier work probing the inner structure of the disk \citep{lagrange_narrow_2016}, \cite{crotts_deep_2021} conducted a deep polarimetric study and found that the disk is asymmetrical in surface brightness and structure. An inclined ring model, fit to the inner disk spine (location of peak surface brightness), suggests that the inner disk may be eccentric ($\gtrsim0.16$). More recent ALMA observation has not confirmed this eccentricity \citep{fehr_millimeter_2022}. Measurements of the vertical height of the inner disk also suggest that it is relatively flat, with an upper limit of a $15.6$ au vertical FWHM \citep{crotts_deep_2021}.

Observations of both the companion and the disk allow us to place significant constraints upon the dynamical history of the system. Previous work has probed the hypothetical dynamical history of the system, including that done by \cite{jilkova_debris_2015}, \cite{nesvold_hd_2017}, and \cite{rodet_origin_2017}. In particular, \cite{rodet_origin_2017} investigated the possibility that HD106906b originally formed within a protoplanetary disk near the central binary. In this scenario, the companion migrated inward, encountered an unstable mean-motion resonance with the binary, and was ejected into a high-eccentricity orbit. This model then requires an external stellar perturber to raise the pericenter of the companion into a stable region. \cite{de_rosa_near-coplanar_2019} searched for potential stellar perturbers consistent with this scenario and discovered two candidates within the currently known members of Sco-Cen that could have had a dynamically important close encounter with HD106906 within the last 15 Myr. However, \cite{rodet_odea_2019} showed that a closest approach distance ($D_{CA}$) of $<0.05$ pc is needed to modify the orbits of either the companion or the disk of HD106906 and the two candidates identified by \cite{de_rosa_near-coplanar_2019} likely had a $D_{CA}>0.5$ pc. Therefore, the asymmetries observed in the debris disk of HD106906 are likely due to perturbations from the companion \citep{jilkova_debris_2015,rodet_origin_2017,nesvold_hd_2017,nguyen_first_2021}.

It is also possible that the companion formed in situ with HD106906 at its current separation and inclination, though this would require a nebula as large as $1000$ au in radius \citep{maury_characterizing_2019}. Gravitational instability in a turbulent environment can help explain the misalignment between the spin axis of the companion, its orbit normal, and the orbit normal of the debris disk \citep{bryan_obliquity_2021}. The companion could be formed in a gravito-turbulent disk around a proto-star \citep{bryan_obliquity_2020,jennings_primordial_2021} or from a portion of a self-gravitating turbulent cloud that fragmented into a binary and companion \citep{bate_formation_2002,bate_stellar_2009,bate_diversity_2018}. However, as we will later show, if the companion formed on an orbit inclined relative to the plane of the disk, long-term torquing of the disk by the gravitational influence of the companion would have long since warped and puffed the structure of the disk.

What additional mechanism can lead to the formation and the misalignment of the companion? A recent survey of free-floating planets in the Upper Scorpius subgroup (USCO), also contained within the Sco-Cen association, suggest that the relative abundance of these objects may be higher than previously thought \citep{miret-roig_rich_2021}. This opens up the possibility that close encounter flybys with these objects, which may range up to brown dwarf mass, can play a significant role in sculpting nearby planetary systems even as they escape detection due to their typically low brightness. A close encounter between one of these free-floating planets and an object with similar mass could excite the eccentricity and inclination of the surviving object while leaving the overall structure of a surrounding debris disk relatively unaffected.

Therefore, we present here an alternative formation scenario for HD106906. The system formed with a large planetary mass companion in the plane of its debris disk. Then, a recent ($\sim1-5$ Myr) close encounter between the companion and a large free-floating planet produced the observed eccentric and inclined orbit of HD106906b while leaving the structure of the disk largely unperturbed. Further interaction between the surviving companion and the debris disk excited the eccentricity and aligned the orbits of the debris disk particles giving rise to the observed disk eccentricity that we see today. Measurements of the flatness of the disk allows us to constrain when this close encounter occurred. As the system continues to evolve, the disk will continue to warp and puff so that the structure will no longer be flat.

In Section \ref{sec:vert_warp}, we carry out numerical simulations to explore the interaction between the eccentric and inclined companion and the debris disk and measure the warping of the disk. This allows us to roughly constrain when the companion was placed in an inclined orbit. In Section \ref{sec:smack} we produce simulated observational images using the software package SMACK \citep{nesvold_smack_2013} in order to compare the end result of our formation scenario to real observed images of the system. In Section \ref{sec:scattering} we model a close encounter between the system and a free-floating planet to explain the current orbit of HD106906b. Finally, in Section \ref{sec:conc} we present our conclusions.

\section{Vertical Warping due to the Companion}\label{sec:vert_warp}
\subsection{Illustration on the Puffing of the Disk}
To test the effects of an inclined companion on the debris disk surrounding HD106906, we modeled the system using a collisionless \textit{N}-body simulation. For our simulations we used the Bulirsch-Stoer integrator in \textsc{Mercury} \citep{chambers_mercury_2012}.

We modeled the binary at the heart of HD106906 in two different ways. In our first model the binary is two individual stars with the primary mass 1.37 $\mathrm{M_{\odot}}$ and the secondary mass 1.34 $\mathrm{M_{\odot}}$. The period of the binary is set to be 49.233 days with an eccentricity of 0.669 based upon results from \cite{de_rosa_near-coplanar_2019}. When we model the binary as two separate stars, the time-step of our simulation is required to be on the order of $\sim2$ days in order to accurately recreate the binary motion. Such a short time-step quickly becomes computationally expensive if we wish to test the dynamics of the system on the timescale of Myrs. Thus we are motivated to find another way to model the dynamics that may arise from the binary in a more efficient way. To this end, our second model of the binary is a central mass with 2.71 $\mathrm{M_{\odot}}$, the combined mass of the binary. We approximate the secular effects of the binary by including a J2 potential around this central mass. To construct our J2 potential we assume the binary has a separation of 0.58 au and is co-planar with the disk. The magnitude of this J2 potential is set to reproduce an equivalent nodal precession in external test particles that would occur around a physical binary.

For each simulation we placed a companion in orbit around the system with 11 $\mathrm{M_{Jup}}$. The semi-major axis was chosen to be $850$au, eccentricity was selected to be 0.4, and the inclination of the companion's orbit relative to the plane of the system was chosen to be $40^\circ$. These values were chosen to be similar to the median values that best fit observation from \cite{nguyen_first_2021}. The argument of pericenter of the companion is set to $0^\circ$ for simplicity since the argument of pericenter of HD106906b is estimated to be close to the disk plane. In the case where we model the binary as two separate stars we chose the longitude of ascending node of the companion such that its orbit is aligned with that of the binary. The companion is initially placed at its apocenter at the start of the simulations. We note that the choice of the companion's longitude of ascending node and its initial phase in its orbit is somewhat arbitrary and does not significantly affect our outcomes.

We then populated the system with a thin disk of $1000$ test particles with a uniform distribution of semi-major axis ranging from $50$ au to $125$ au similar to the inner ring model used to compare with observation in \cite{crotts_deep_2021}. We assume the initial orbits of the particles are near circular and near co-planar, with eccentricities randomly uniformly selected from $0$ to $0.01$ and inclinations randomly uniformly selected from $0$ to $0.1$ degrees. We note that the initial eccentricity and inclination of the test particles do not affect our results qualitatively. The argument of pericenter, longitude of ascending node, and initial phase of each test particle were chosen at random.

Observations of the inner disk are dominated by the surface brightness of dust particles. These dust particles are produced when planetesimals within the disk collide which is more likely to occur at shorter disk radii. These dust particles are subject to additional forces from the radiation of the central stars, including radiation pressure and the Poynting-Robertson (P-R) drag force \citep{wyatt_how_1999} due to their small size. The radiation pressure acts radially on the dust particles and has a similar dynamical effect of reducing the central mass that the dust particle orbits. The P-R drag force acts tangentially on the dust particles and results in an evolutionary decrease in the semi-major axis and eccentricity of the dust particles. However, the magnitude of the the P-R drag force is typically very small so we neglect this effect here. On the other hand, the effects of radiation pressure are likely significant.

The dust particles that result from collisions likely have a similar initial velocity vector as the parent objects that collided and produced them. The smallest of these particles are ejected from the system as the radiation pressure exceeds the gravitational force from the central mass. However, slightly larger dust grains continue to be bound to the central mass and continue on wider orbits that are more eccentric than the orbits of their parent objects. External perturbations of these dust particles may also be more effective since, dynamically, the central mass is functionally reduced from their perspective.

Thus, for each model of the central binary, we created one simulation which modeled the effects of radiation pressure and one which neglected these effects. In the simulations where we modeled the effects of radiation pressure,  each dust particle has a different $\beta$ value (where $\beta=F_{rad}/F_{grav}$ is the ratio of radiative and gravitational forces acting on a dust grain). The $\beta$ values were randomly chosen from a power-law distribution with an index 3/2. The maximum possible value for $\beta$ is set by the parameters of the parent body's orbit,
\begin{equation}
    \beta_{max}=\frac{1-e_p^2}{2(1+e_p\cos{f_p})} \ ,
\end{equation}
where $e_p$ and $f_p$ are the original test particle's eccentricity and true anomaly, respectively. Smaller particles with a $\beta$ value larger than $\beta_{max}$ will have initially hyperbolic trajectories and will be quickly removed from the system, so we exclude them. This selection of $\beta$ values leads to an expectation value of $\langle\beta\rangle\approx0.34$.

This $\beta$-distribution is related to the assumed size distribution of the dust grains resulting from a collision. By assigning different $\beta$ values to each particle we are in effect simulating a disk composed of dust particles with different sizes. If the size distribution derives from a standard collisional cascade and obeys a Dohnanyi distribution $dN/ds\propto s^{-7/2}$, where $s$ is the particle size, then $dN/d\beta\propto\beta^{3/2}$. This is under the assumption that dust particles present geometric cross sections to radiation pressure ($\beta\propto 1/s$), following \cite{wyatt_how_1999}, \cite{lee_primer_2016}, and \cite{nesvold_hd_2017}. If we roughly estimate the density of the dust particles to be $1$g $\mathrm{cm}^{-3}$, our expectation value of $\beta$ corresponds to an average dust size of $\sim60$ $\mathrm{\mu}$m (using Eqn. 18 of \citealt{wyatt_how_1999}) so this assumption should hold. This corresponds to a range of dust grain sizes from $\sim37$ $\mathrm{\mu}$m to $\sim5,500$ $\mathrm{\mu}$m. However, due to the $\beta$-distribution we are selecting from, mm-size particles should be very rare in our simulations. We estimate that less than $0.01\%$ of particles in our simulations are larger than 1 mm. We note that the peak wavelengths observed by GPI in \cite{crotts_deep_2021} are of the order $\sim1$ $\mathrm{\mu}$m so we are simulating grain sizes that are comparable to what would be probed with observation.

We then simulate the evolution of the dust particles under the gravitational influence of the central mass and companion, and a radiation pressure force proportional to the force of gravity from the central mass:
\begin{equation}
    \vec{F}_{rad}=-\beta\vec{F}_{grav} \ .
\end{equation}
These simulations are then evolved to $5$Myr.

In summary, we model the central binary in two different ways: first as two separate stars, then second as a central mass with an equivalent J2 potential. For each of these two models we simulate with and without the effects of radiation pressure from the stars on the test particles. The orbital parameters of the external companion are identical in each simulation.

As the systems evolve, we measure the eccentricity of the disk and its vertical warping over time. The eccentricity of the disk is measured by calculating the offset of the center of the disk (average position of all the test particles) relative to the location of the central mass. The eccentricity is then simply the offset distance divided by the disk semi-major axis.

To study more closely the evolution of the eccentricity of the inner disk, we included another series of simulations with the effects of radiation pressure in which we fully populate the inner region of the system with test particles. These particles begin with a uniform distribution of semi-major axis ranging from $10$ au to $20$ au, eccentricities randomly uniformly selected from $0$ to $0.01$ and inclinations randomly uniformly selected from $0$ to $0.1$ degrees. Radiation pressure quickly increases the semi-major axes of these particles so that the inner region of the system is well populated with an average distance of $100$ au, creating a similar size disk as the simulations which do not include radiation pressure.

The vertical warping of the debris disk is measured by calculating the vertical FWHM of the particles' positions above and below the plane of the disk. At each time-step of the simulation, the plane of the disk is defined as being perpendicular to the average specific angular momentum vector of the disk. Redefining the plane of the disk at each time step allows us to measure the vertical height of the disk even if it is warped away from its original plane. We then discard particles with a vertical distance $\geq3\sigma$ from the mean and $>125$ au as we consider such outliers no longer part of the inner disk bulk. The remaining particles roughly follow a Gaussian distribution above and below the disk plane, and so we simply measure the vertical FWHM of the distribution. Due to the sensitivity of this measurement to locations of individual particles at each time-step we take the moving average of the vertical FWHM over time. 

Calculating the vertical FWHM of the particles' positions above and below the disk plane allows us to compare our results with observations of the vertical height of the disk surrounding HD106906 made by \cite{crotts_deep_2021}. Specifically, \cite{crotts_deep_2021} measured the surface brightness along vertical cross-sections of the inner disk, fit a Gaussian to each vertical brightness profile, and then calculated the FWHM at each cross-section. Their results suggest an intrinsic vertical FWHM of 15.6 au. It is important to note that the vertical FWHM that we are calculating is based upon particle positions, while the vertical FWHM that \cite{crotts_deep_2021} measure is based upon brightness of pixels in their observations, so the comparison is not perfect and should only be taken as a rough estimate. There is also some degeneracy between the inclination of the debris disk and its warp in the observational results. However, our view of HD106906 is nearly edge-on with a reported inclination of $\sim5^\circ$ relative to our line of sight \citep{crotts_deep_2021,lagrange_narrow_2016,kalas_direct_2015}. Therefore, we use the vertical FWHM of \cite{crotts_deep_2021} only as an upper limit on the puffiness of the disk.

\begin{figure*}[htb]
\centering
    \includegraphics[width=0.97\textwidth]{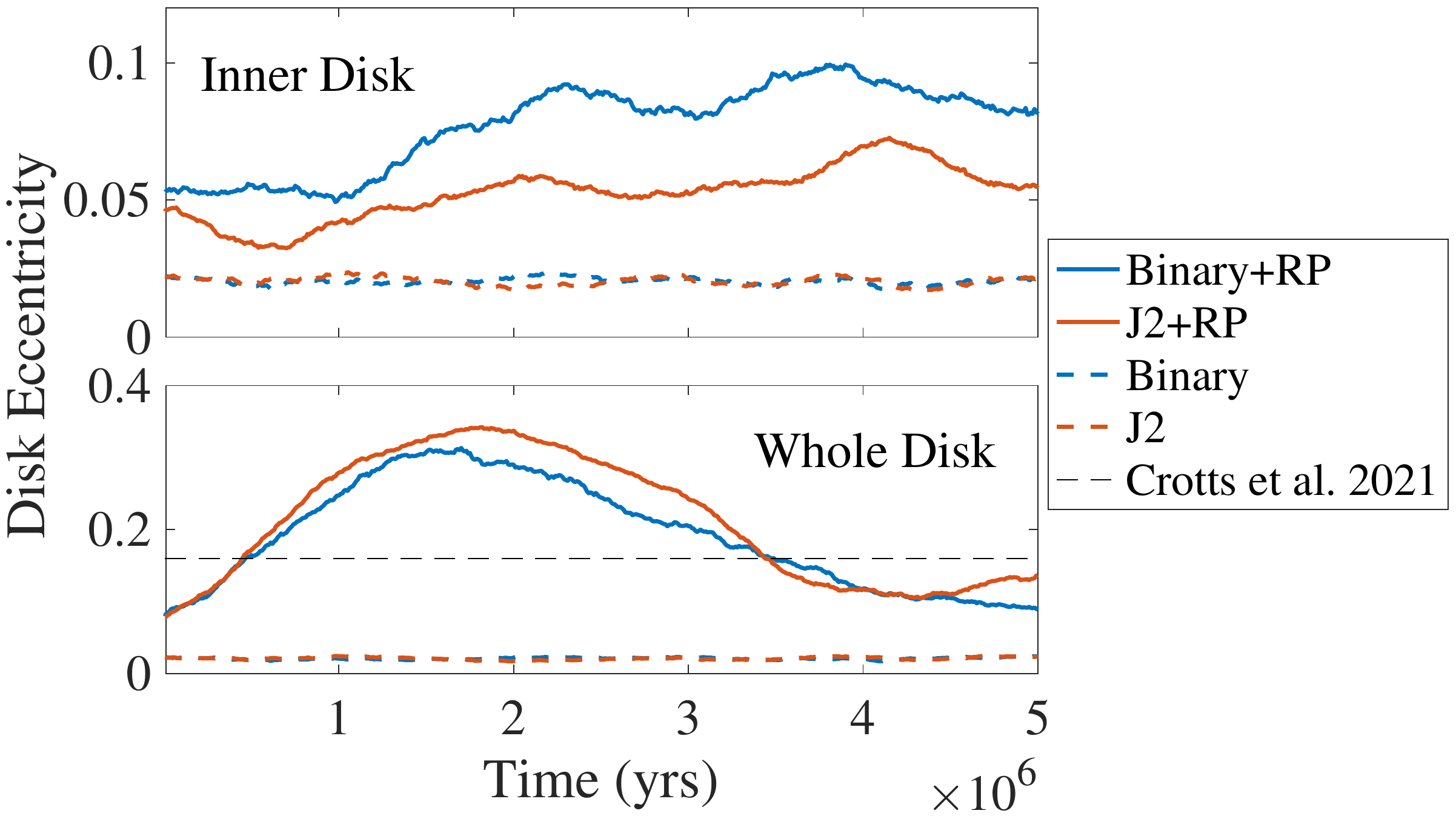}
    \caption{Disk eccentricity under the influence of an 11 $\mathrm{M_{Jup}}$ mass companion with best fit orbital parameters from \cite{nguyen_first_2021}. \textbf{Top Panel:} Focusing on inner disk particles. Simulations with radiation pressure have their test particles starting closer to the central mass. \textbf{Bottom Panel:} Considering particles in the whole disk. In both panels the solid lines show the simulations with the effects of radiation pressure, the dashed lines without. The blue lines correspond to simulations where we modeled the central mass as two separate stars. The orange lines correspond to simulations where the central mass is modeled as a single star with an equivalent J2 potential. The disk observational constraints of \cite{crotts_deep_2021} (e$\gtrsim0.16$) are shown with the dashed black line in the \textbf{bottom panel}.}
    \label{fig:disk_ecc}
\end{figure*}

The top panel of Figure \ref{fig:disk_ecc} shows the disk eccentricity over time. Simulations with radiation pressure have their test particles starting closer to the central mass. In the simulations without radiation pressure the disk eccentricity remains low and roughly constant at around 0.02 for both models of the central mass. For simulations with radiation pressure the disk eccentricity is higher for both models of the central mass. The J2+RP model peaks at $\sim0.07$, and the Binary+RP model peaks at $\sim0.1$. In all cases, even if the eccentricities of the individual particle orbits are growing, the nodal precession of the orbits due to the effects of the binary can cause the orbits to become misaligned.

However, in the bottom panel of Figure \ref{fig:disk_ecc}, if we extend our analysis to include particles that lie beyond $125$ au where the timescale of the nodal precession is much longer, we find that the external companion is able to excite a cohesive disk eccentricity in the simulations that include the effects of radiation pressure. In these simulations the disk eccentricity peaks near 0.3 just before $2$ Myr. After this time the nodal precession of the orbits again cause a misalignment and the disk eccentricity drops. Again, without radiation pressure, the disk eccentricity remains low.

The eccentricity of the inner disk of HD106906 reported in \cite{crotts_deep_2021} is $\sim0.16\pm0.02$. Our simulation of the inner disk with the Binary+RP model is able to reach a similar moderate value at its peak of $\sim0.1$, though our other simulations cannot excite a disk eccentricity of similar magnitude. \cite{farhat_case_2022} does point out that there remains significant uncertainty in the observed orbit of HD106906b and this can lead to a wide range of disk eccentricities. Based upon ALMA observations, \cite{fehr_millimeter_2022} does not measure a noticeable inner disk eccentricity. However, since their observations are probing larger dust grains that are not subject to the effects of radiation pressure, our results from simulations without radiation pressure are consistent with their measurement.

\begin{figure*}[htb]
\centering
    \includegraphics[width=0.97\textwidth]{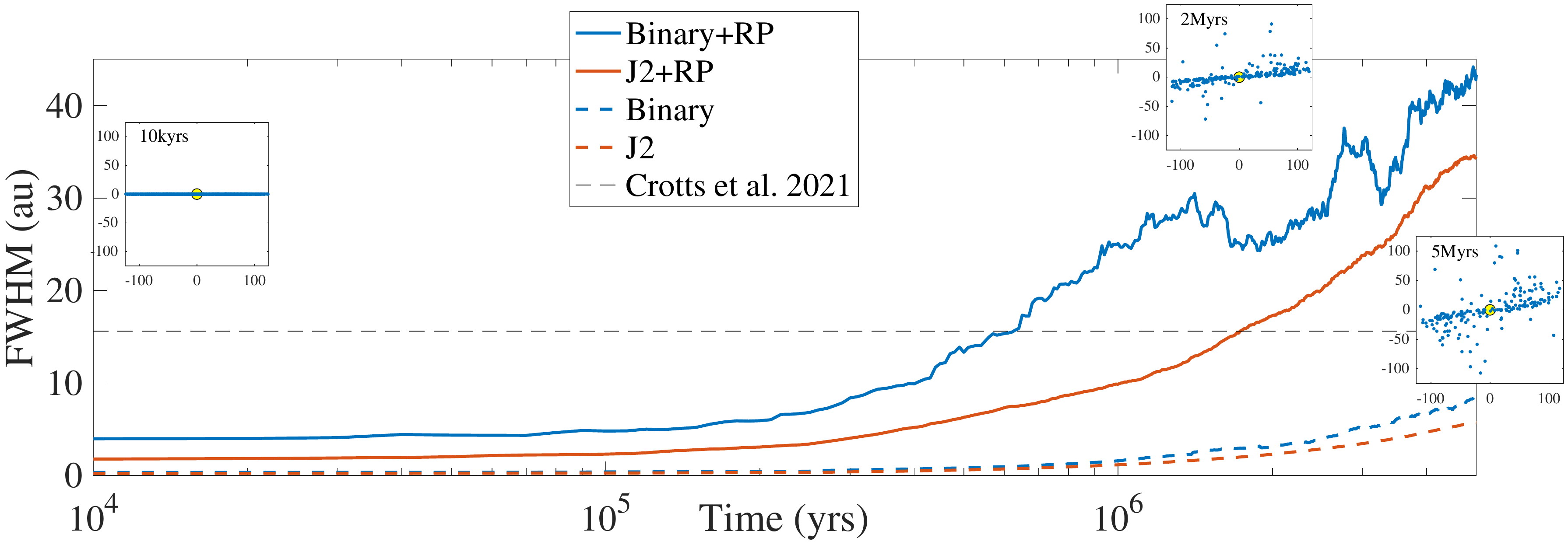}
    \caption{Vertical FWHM of debris disk particles' vertical position under the influence of an 11 $\mathrm{M_{Jup}}$ mass companion with best fit orbital parameters from \cite{nguyen_first_2021}. The solid lines show the simulations with the effects of radiation pressure, the dashed lines without. The blue lines correspond to simulations where we modeled the central mass as two separate stars. The orange lines correspond to simulations where the central mass is modeled as a single star with an equivalent J2 potential. The disk observational constraints of \cite{crotts_deep_2021} (FWHM$<15.6$ au) are shown with the dashed black line. Also shown in the subplots is a side-on view of the disk (from the J2+RP model) at three different times of interest. The units of these subplots are au. When the central mass is modeled as two separate binaries the puffing of the disk is accelerated. Including radiation pressure also causes the disk to puff up more rapidly such that it exceeds observational constraints by $<2$Myr.}
    \label{fig:vd_snapshots}
\end{figure*}

For the remainder of our paper, we mostly focus on the well constrained flatness of the disk by measuring the vertical FWHM of the debris disk particles. Figure \ref{fig:vd_snapshots} shows the vertical FWHM of debris disk particles' positions over time. We can see that, when we include the effects of radiation pressure, the influence of the eccentric and inclined companion causes the disk to vertically warp relatively rapidly. The first time step which the vertical FWHM of the disk exceeds 15.6 au is just over $0.5$Myr in the case of the binary model, and just over $1.75$Myr in the case of the J2 model. By $>2$Myr, the structure of the disk is severely warped and the flatness of the disk is no longer recognizable when we include radiation pressure. The effect of the radiation pressure on the distribution of dust particles can be quite significant, since the effect of radiation pressure on dust particles can be seen as equivalent to reducing the mass of the central binary by a factor of $1-\beta$. The result is that the perturbing influence of an external companion on dust can be more effective than for particles that are large enough to neglect radiation pressure. This enhanced effectiveness of external perturbations is able to overcome inclination damping from nodal precession of the dust orbits due to the secular effects the binary.% since dust particles produced from a colliding parent body can adopt orbits that are much wider and eccentric. For example, if a parent body has a semi-major axis of 50 au and eccentricity of 0.05, dust grains that are small enough that their $\beta$ value approaches 0.4 will begin to orbit the central mass with a semi-major axis nearly quadruple in size and with an eccentricity of 0.75.

It is important to note that vertical warping and puffing of the disk is faster in the binary model, with or without the effects of radiation pressure. Therefore, when we use the J2 model to estimate the timescale of the vertical warping and puffing of the disk in subsequent simulations to save computation time, the estimate should be taken as an upper bound since a real binary is typically more disruptive.

These results are consistent with the findings of \cite{jilkova_debris_2015, nesvold_hd_2017} which found that a disk under the influence of a moderately inclined companion ($\sim30^\circ$) can vertically warp the disk so that it is no longer consistent with observations of HD106906. In fact, these earlier results were previously used to attempt to dynamically constrain the inclination of the companion's orbit and suggest a nearly co-planar configuration for the system. However, since the mutual inclination of the companion has recently been observationally constrained to be $>35^\circ$ \citep{nguyen_first_2021}, we suggest that the companion had its inclination excited relatively recently ($\lesssim5$ Myr).

\subsection{Dependence on the Companion Eccentricity and Inclination}\label{sec:ecc_and_inc}
We designed another series of simulations which allows us explore the interaction between the disk and companion by widely varying the eccentricity and inclination of the companion's orbit. We tested 72 different orbits of the companion. In each simulation the semi-major axis of the companion is $850$ au. The companion's orbital eccentricities ranged from $0.1$ to $0.9$ in intervals of $0.1$. The companion's orbital inclination ranged from $10^\circ$ to $80^\circ$ in intervals of $10^\circ$. The argument of pericenter of the companion is again set to $0^\circ$. The central binary is modeled using a central mass and J2 potential. The debris disk is again modeled by populating the system with a thin disk of $1000$ test particles with a uniform distribution of semi-major axis ranging from $50$ au to $125$ au. These simulations include the effects of radiation pressure. All companions begin the simulation at their apocenter. The systems are then integrated for $5$ Myr. At each time step the vertical FWHM of the particles' positions above and below the plane of the disk is measured. 

\begin{figure*}[htb]
\centering
    \includegraphics[width=0.97\textwidth]{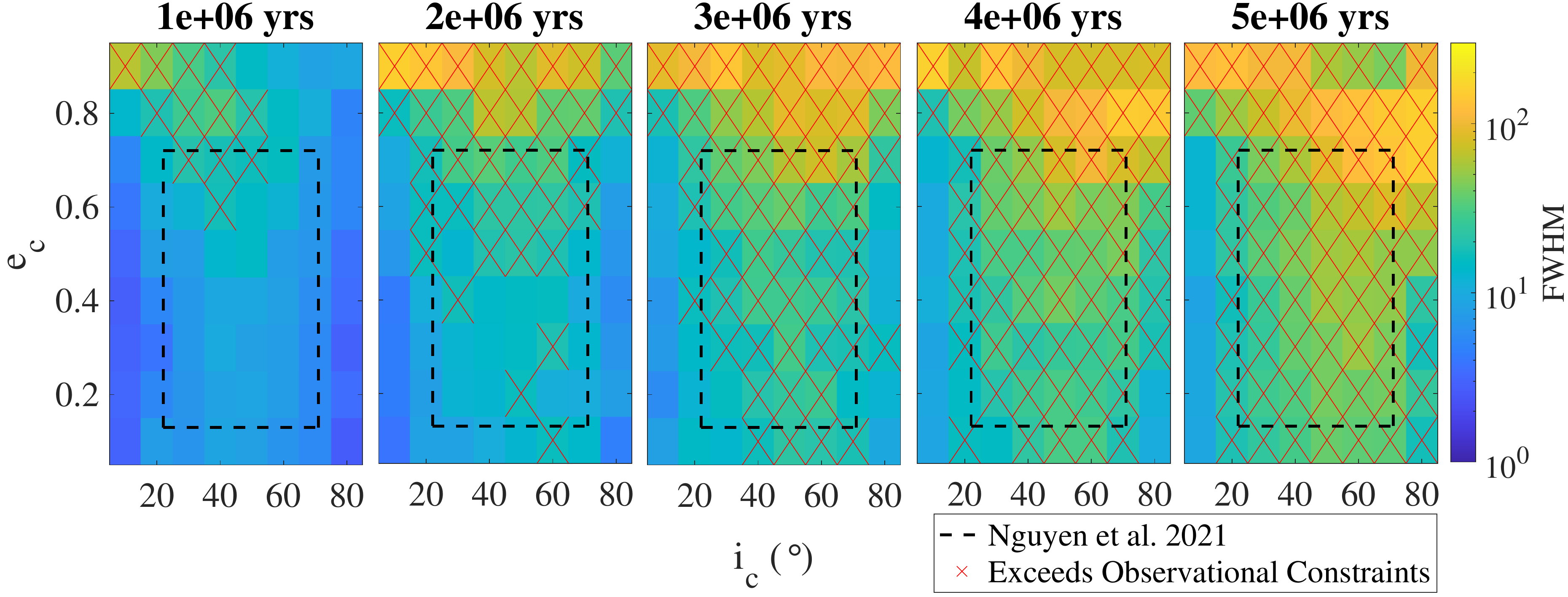}
    \caption{Vertical FWHM of disk particle positions under the influence of an exterior companion at different time steps. The vertical FWHM is shown by the color of each pixel. Each panel is a different time step. Within each panel the x axis denotes the companion inclination ($\mathrm{i_c}$) and the y axis denotes the companion eccentricity ($\mathrm{e_c}$). The observational constraints of HD106906b reported by \cite{nguyen_first_2021} is shown by the dashed black line. Pixels are covered in a red ``X" if the vertical FWHM of disk particle positions exceeds 15.6 au. Systems with a highly eccentric companion have their disk vertical FWHM excited most rapidly. Dependence of vertical FWHM on companion inclination is weak. By $4$ Myr there are no companion configurations within the observational constraints of \cite{nguyen_first_2021} which preserve the flatness of the disk.}
    \label{fig:disk_evo_params}
\end{figure*}

Figure \ref{fig:disk_evo_params} shows the results from these simulations. Companions with a high eccentricity excite the disk vertical FWHM most rapidly. For the most eccentric configurations ($\mathrm{e_c}=0.9$) the pericenter of the companion lies within the outer edge of the disk ($125$ au) and so multiple short-term close encounter scattering events puff the disk very rapidly ($\lesssim1$ Myr for $\mathrm{i_c}\leq40^\circ$).

For more moderate companion eccentricities, the gravitational perturbations from the companion on the disk tend to excite particles' eccentricities and inclinations \citep[e.g.,][]{naoz_secular_2013}. This process highly depends on the particle's separation and thus naturally results in warping through the Eccentric Kozai-Lidov (EKL) mechanism \citep[e.g.,][]{nesvold_circumstellar_2016,nesvold_hd_2017}. The timescale of EKL oscillations ($\tau_{EKL}$) depends upon the eccentricity of the outer companion by a factor $(1-e_c^2)^{3/2}$. Thus, systems with a larger companion eccentricity have a shorter $\tau_{EKL}$ and warp to become puffy more rapidly.

Dependence of vertical FWHM upon the companion inclination is weak. We do note that the lowest inclination companions ($i_c\lesssim20^\circ$) can preserve the disk flatness for long period of time except in the cases of extreme companion eccentricity ($e_c\gtrsim0.8$). However, these low inclination configurations are unlikely due to the observational constraints of \cite{nguyen_first_2021}.

By $4$ Myr there are no companion configurations within the observational constraints of \cite{nguyen_first_2021} which preserve the flatness of the disk. These results underscore how rapidly a disk can become warped in the presence of an inclined companion.

\subsection{Dependence on the Companion Semi-Major Axis}
We designed a new series of simulations in which we restrict our focus to the observational constraints of the companion. In these simulations we vary not only the inclination and eccentricity of the companion, but also the semi-major axis and argument of pericenter. With these new simulations we can test the vertical warping of the disk with many different companion orbits and orientations.

We tested $1000$ different orbits for the companion. The semi-major axes were randomly uniformly selected on the range $590$-$1410$ au and the eccentricities were uniformly selected on the range $0.13$-$0.72$. These ranges were chosen as they are $\pm1\sigma$ the median value of the best fit companion orbital parameters from \cite{nguyen_first_2021}. The orientations of the orbits (argument of pericenter and mutual inclination) were also randomly uniformly selected from the $\pm1\sigma$ ranges from the two families of orbital parameters in \cite{nguyen_first_2021}. For the first half of the simulations the inclination range (relative to the original disk plane) was $22^\circ-63^\circ$ and the argument of pericenter range was $270^\circ-57^\circ$. For the second half of the simulations the inclination range was $30^\circ-71^\circ$ and the argument of pericenter range was $89^\circ-237^\circ$. We note that the posterior distributions of likely orbital parameters of the companion from observation means that some combinations of orbits and orientations are more likely that others. However, the goal of our uniform sampling of orbits is to fully explore the parameter space of possible orbits and test the dynamics of the companion-disk interaction.

The central binary is modeled using a central mass and J2 potential. The debris disk is modeled by populating the system with a thin disk of $1000$ test particles with a uniform distribution of semi-major axis ranging from $50$ au to $125$ au subject to radiation pressure. The systems are then integrated for $10$ Myr. At the end of each simulation we calculated the vertical warping timescale ($\tau_\mathrm{VW}$) for each orbit. The vertical warping timescale is defined as the first time step at which the vertical FWHM of the positions of disk particles exceeds $15.6$ au. 

For reference, we also calculate the Eccentric Kozai-Lidov (EKL) oscillation timescale ($\tau_\mathrm{EKL}$) for a test particle at $100$ au using the minimum and maximum companion semi-major axes and eccentricities. This gives us a benchmark timescale with which to qualitatively compare the vertical warping. However, as mentioned above, the vertical warping quickly develops and thus only a fraction of the Eccentric Lidov-Kozai timescale is required. The timescale for the restricted three-body problem is approximately given by \citep[e.g.,][]{kinoshita_analytical_1999,naoz_eccentric_2016}:
\begin{equation}
    \tau_{\rm VW}\sim\alpha \tau_{\rm EKL}  \sim \alpha \frac{P_2^2}{P_1}\frac{m_1+m_3}{m_3}\left(1-e_3^2\right)^{3/2}
\end{equation}
where $P_2$, $e_3$, and $m_3$ are the period, eccentricity, and mass of the outer orbit of the companion, respectively. $m_1$  is the mass of the central binary and the orbital period of the inner disk particle is $P_1$. $\alpha$ is a fractional coefficient. Particles that orbit at less than $100$ au have a longer EKL timescale, and particles that orbit at greater than $100$ au have a shorter EKL timescale.

%It is important to note that the vertical warping timescale is shorter than the full EKL cycle, since the inclination only needs to increase slightly to break the observational constraints (e.g., a FWHM of $15.6$ au). Our simulations show that average disk inclination only needs to be excited to a few percent of the maximum expected value from a full EKL cycle in order to break these constraints.

\begin{figure}[htb]
\centering
    \includegraphics[width=0.47\textwidth]{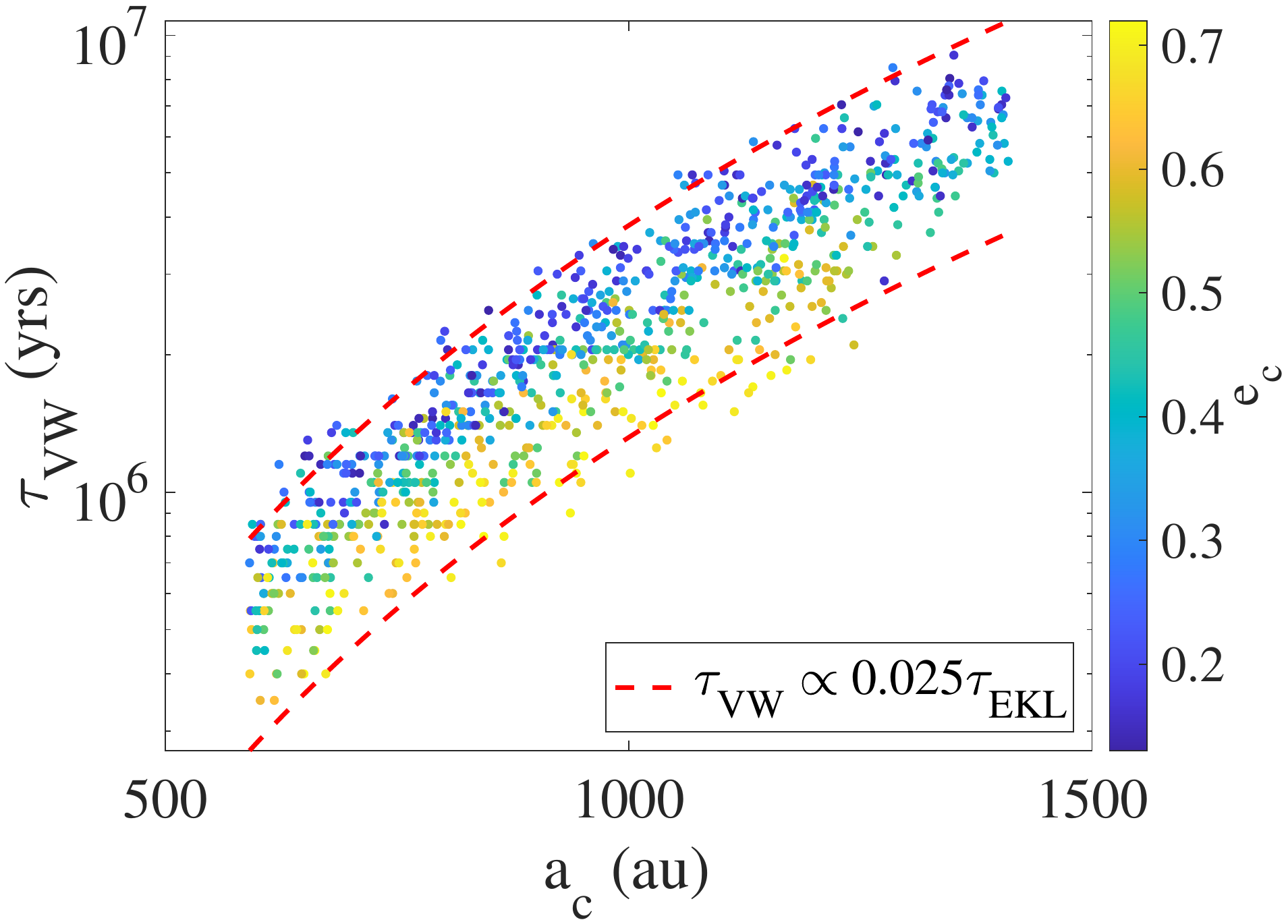}
    \caption{Vertical warping timescale of the debris disk for companions with different orbital elements selected from best fit parameters from \cite{nguyen_first_2021}. Here the timescale is displayed as it depends upon companion semi-major axis. Points are colored by the eccentricity of the companion. The dashed red lines show the bounds of the EKL timescale for a disk particle at 100 au (reduced to $2.5\%$ in magnitude) where the bottom and top lines correspond to the maximum and minimum companion eccentricities we tested respectively. The companions with a smaller semi-major axis tend to warp the disk on shorter timescales.}
    \label{fig:vd_timescale}
\end{figure}

We plot the vertical warping timescale of each system in Figure \ref{fig:vd_timescale}. We find that systems that have a companion with a smaller semi-major axis have a shorter vertical warping timescale. Reinforcing our results from Section \ref{sec:ecc_and_inc}, we again find that systems with a more eccentric companion tend to warp this disk on smaller timescales. We also again find little dependence upon companion inclination and companion argument of pericenter. The warping of the disk is driven primarily through the secular inclination excitation during the beginning of the first EKL cycle of the disk particles. We find that the vertical warping timescale is typically $\sim2.5\%$ of the EKL timescale.

%there are no more cases in these simulations where the companion encroaches into the inner disk, no iEKL discussion

In many of these cases the disk becomes warped after only a few $100$s of kyr, and typically on the order of $\sim1$ Myr. We should note that this vertical warping timescale is an upper estimate due to the way we modeled the central binary. The growth of the vertical FWHM of the disk is likely faster if we model the central mass as two separate stars (see Figure \ref{fig:vd_snapshots}).

Only about $10\%$ of systems have a vertical warping timescale of longer than $5$ Myr. These systems typically have a companion that is on an extremely wide orbit with a semi-major axis greater than $1000$ au. While these configurations are allowed, the correlated semi-major axis and eccentricity distributions from the observational constraints of \cite{nguyen_first_2021} show that they are very unlikely.

To summarize, these findings suggest that the present configuration of the system, HD106906, cannot have survived for longer that $\sim5$ Myr without causing significant vertical warping to the disk. Since the age of the system is $13\pm2$ Myr \citep{bailey_hd_2014}, some dynamical event must have occurred within its recent history which produced the wide and inclined orbit of HD106906b.

\section{Simulated Observational Images}\label{sec:smack}
It is limiting to compare our \textit{N}-body results to observational images of the inner disk of the system. Our analysis of the \textit{N}-body results are based upon the positions of the test particles. On the other hand, observations are dominated by the surface brightness of dust particles which are produced by collisions between planetesimals within the inner disk.

Therefore, in order to test the validity of the analysis of the \textit{N}-body simulations with regard to their applicability to HD106906, we produce simulated observational images in order to compare the end result of our formation scenario to real observed images of the inner disk of the system. Specifically, we wish to directly compare to the H-band observations (peak $\lambda=1.647$ $\mathrm{\mu}$m) of the inner disk from \cite{crotts_deep_2021}. We chose the H-band observation for comparison since it has the highest signal to noise ratio. To model the dust producing collisions between particles in the disk, we used the software package SMACK \citep{nesvold_smack_2013}, based on the N-body integrator REBOUND \citep{rein_rebound_2012}. SMACK approximates each particle in the integrator as a collection of smaller bodies called a "superparticle" which travel on the same orbit. Each superparticle is approximated as a sphere of some finite radius, and when two superparticles overlap, SMACK models the outcome of the collision. SMACK then conserves the angular momentum and energy of the superparticles by correcting their trajectory after the collision. This allows SMACK to compensate for any energy lost due to fragmentation. Using SMACK we can record dust producing events to generate an accurate initial distribution of dust particles as well as capture any dynamics that may result from collisions within the disk.

To simulate the debris disk surrounding HD106906, we use $10000$ superparticles with semi-major axes randomly uniformly selected on the range $30$-$600$ au. This range of semi-major axes are chosen since the disk is observed to extend out past $500$ au on one side in scattered light images. Collisions between planetesimals are more likely at smaller radii and this initial distribution of planetesimals can recreate an inner disk of dust that is similar to that observed. The eccentricity and inclination of each superparticle is randomly uniformly selected on the range $0$-$0.1$ (degrees in the case of inclination). The longitude of ascending node, argument of pericenter, and mean anomaly for each particle are also randomly uniformly selected to ensure an initially azimuthally symmetric setup.

We then place a companion in orbit around the debris disk with a semi-major axis of $850$ au, an eccentricity of $0.4$, and an inclination of $40^\circ$; similar to the median best fit values of \cite{nguyen_first_2021} and the same set up as our initial simulations. Our previous collisionless N-body simulations in Section \ref{sec:vert_warp} showed that a companion with these orbital parameters is able to excite the inner disk vertical FWHM to $>15.6$ au by $2$ Myr. The argument of pericenter of the companion is set to $0^\circ$ for simplicity since the argument of pericenter of HD106906b is estimated to be close to the disk plane. Due to the initially azimuthally symmetric setup of the disk, we can arbitrarily set the longitude of ascending node of our companion to be $0^\circ$. The companion begins the simulation at its apocenter. We then evolve the system for $10$ Myr.

For each dust production event that SMACK records, we generate $1000$ dust orbits. We assume that the dust particles share the same initial velocity of the parent objects that collided and subsequently have their initial orbits modified by the effects of radiation pressure. Each dust particle produced has a different $\beta$ value which are assigned from the same power-law distribution $dN/d\beta\propto\beta^{3/2}$ described in Section \ref{sec:vert_warp}. This implies the same assumed size distribution of the dust particles $dN/ds\propto s^{-7/2}$.

We then pass the information of these dust producing events (time, location, and $\beta$ values) into \textsc{Mercury}. We simulate the evolution of the dust particles under the gravitational influence of the central mass and companion, including the effects of radiation pressure. We also include a J2 quadrupole potential around the central mass for the \textsc{Mercury} simulations, in order to model the nodal precession effects of the dust orbits that will arise near the central binary. Note that we showed in Figure \ref{fig:vd_snapshots} that the vertical FWHM of the disk will grow on a longer timescale than if we simulated the central mass as two separate stars.

We neglect the J2 potential for our SMACK simulations which produce our initial dust distribution. We do this because the collisions that produce the dust particles typically take place at low inclinations ($\sim3^\circ$) and thus we believe adding the J2 potential would have a negligible impact on the collisional dynamics.

To construct our simulated observational image, we project the location of each dust particle at different timesteps of the simulation onto a 2D grid with resolution 2 au to create a face-on image. We then simulate the surface brightness of the dust using $\phi(g,\theta)/\beta^2r^2$, where $\phi(g,\theta)$ is the Henyey-Greenstein scattering phase function \citep{henyey_diffuse_1941} with asymmetry parameter $g$, $\theta$ is the angle between the dust grain and the observer's line-of-sight (where the star is the vertex), and $r$ is the distance between the dust particle and the star. Following \cite{lee_primer_2016} and \cite{nesvold_hd_2017}, we chose $g=0.5$. Finally, we rotate our viewing angle of the system so that the estimated pericenter of the companion is oriented on the same side as observation, and we tilt the disk so that it is inclined $5^\circ$ relative to our viewing angle.

This analysis implicitly assumes that the emission is optically thin, an assumption also held in previous observational studies of the system \citep[e.g.,][]{lagrange_narrow_2016}. The polarization data from \cite{crotts_deep_2021} is also consistent with an optically thin debris disk causing single scatterings.

We then measure the intrinsic FWHM of the disk in the exact same way described in \cite{crotts_deep_2021}. Specifically, we measured the surface brightness along vertical cross-sections of the inner disk (after we rotate our image so that the disk’s major axis is horizontal), fit a Gaussian to each vertical brightness profile, and then calculated the FWHM at each cross-section. The intrinsic FWHM is the weighted mean FWHM.

\begin{figure*}[htb]
\centering
    \includegraphics[width=0.97\textwidth]{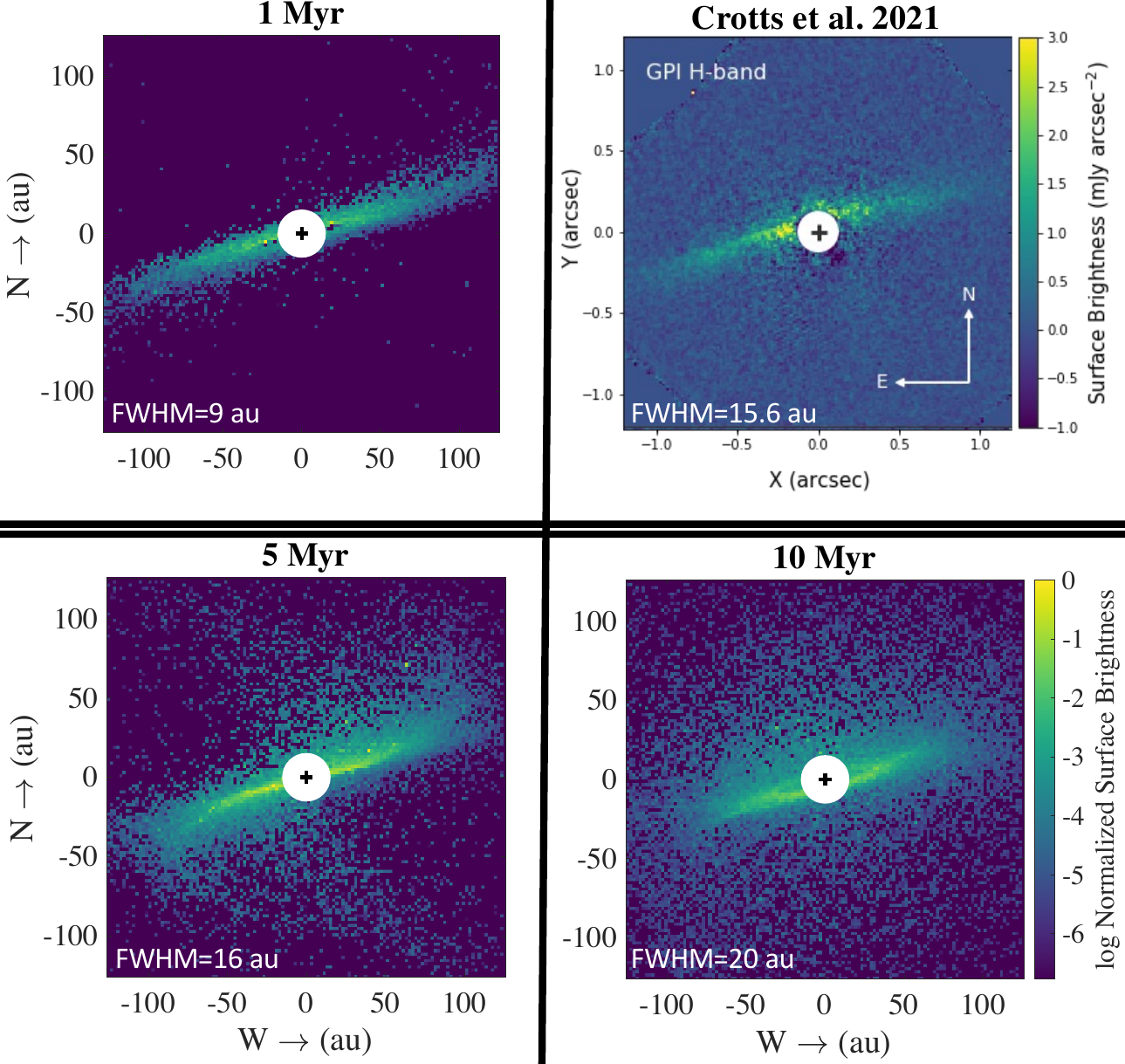}
    \caption{Simulated vs real observational images of the inner disk of HD106906. The real observational image (top-right) is taken directly from Fig. 1 of \cite{crotts_deep_2021} (reproduced by permission of the AAS). The simulated observational images (top-left, and bottom row) have a fake coronagraph mask to facilitate comparison. The simulated image at $1$ Myr share many qualitative similarities with the real image, including a peak in surface brightness on the east side of the disk and a slight vertical dispersion on the west side. At this time the FWHM of the disk is below observational constraints. By $5$ and $10$ Myr, the FWHM of the disk exceeds observational constraints. $1$ arcsec is $\sim100$ au.}
    \label{fig:smack_images}
\end{figure*}

Figure \ref{fig:smack_images} shows the results of our simulated observational images compared to the H-band observation of HD106906 from \cite{crotts_deep_2021}. We find that our simulated observational image of the system after $1$ Myr of evolution compares favorably with observation. Qualitatively, the peak brightness of the simulated image is on the east side of the disk, similar to the location of the peak brightness of the real image. We also find that at $1$ Myr the flatness of the disk lies within the observational constraints of \cite{crotts_deep_2021}. Specifically, we measure the vertical FWHM of the disk at this time to be $\sim9$ au.

By $5$ Myr, external torquing by the companion, made more effective by radiation pressure, has caused some inclination excitation of the dust particles and the disk has become puffy. The vertical FWHM of the disk at this time is $\sim16$ au and exceeds observational constraints. We also notice some slight warping of the disk spine that is qualitatively different to the straighter spine seen in \cite{crotts_deep_2021}.

By $10$ Myr, the puffiness of the disk still exceeds observational constraints with a vertical FWHM of $\sim20$ au. Nodal precession of the dust particle orbits, resulting from the J2 potential of the central binary, has removed the slight warping of the spine that we noticed at $5$ Myr. However, the spine itself still remains puffy.

In these simulated observational images, the dust particles whose inclinations are excited by the external companion eventually have their orbit orientations effectively randomized by the J2 potential of the central binary. Eventually, the resulting structure is a puffy disk with a diffuse cloud of particles above and below the disk plane. We note that noise in the observations might wash out this finer detailed structure making the disk appear flatter than it is in reality.

These results suggest a similar vertical warping timescale than our analysis in Section \ref{sec:vert_warp}. It is clear that by $\sim5$ Myr the influence of the external inclined companion warps the structure of the inner disk so that it is no longer recognizable. This warping process would likely have taken place faster in our simulations with a physical binary at the heart of the system. Depending upon the exact orbital parameters of HD106906b, these findings show that the companion must have had its inclination excited within $\sim1-5$ Myr.

\section{Scattering by Free-Floating Planet}\label{sec:scattering}
In previous sections, we have established that the current configuration of HD106906 cannot survive for long ($\gtrsim5$ Myr) without severely warping the disk. These results suggest that some dynamical event, which resulted in the eccentric and inclined orbit of HD106906b, must have happened relatively recently. In this section we explore the possibility that the wide and inclined orbit of HD106906b was caused by a close encounter with a free-floating planet. A close encounter with a free-floating planet, as opposed to a passing star, could produce a companion on an eccentric and inclined orbit while leaving the structure of the debris disk relatively undisturbed. In addition, if a star is responsible for scattering the companion into an inclined orbit, raising the pericenter of the companion, or if the companion was captured from another system during a stellar flyby event, it is likely that we would be able to identify the flyby star responsible for these scenarios. A recent survey of nearby stellar neighbors of HD106906 by the \textit{Gaia} DR2 catalogue found only two candidate perturbers out of 461 stars analyzed \citep{de_rosa_near-coplanar_2019}. The time of closest approach for both of these candidates is likely $>2$ Myr and the candidates had a median closest approach distance within 1 pc of HD106906. However, \cite{rodet_odea_2019} found that a distance of closest approach of $\lesssim0.05$ pc is required to account for the planet probable misalignment with respect to the disk plane as well as the disk morphology. On the other hand, a free-floating planet could have escaped detection by our observational surveys.

\subsection{Numerical Simulation}
\begin{figure}[htb]
\centering
    \includegraphics[width=0.47\textwidth]{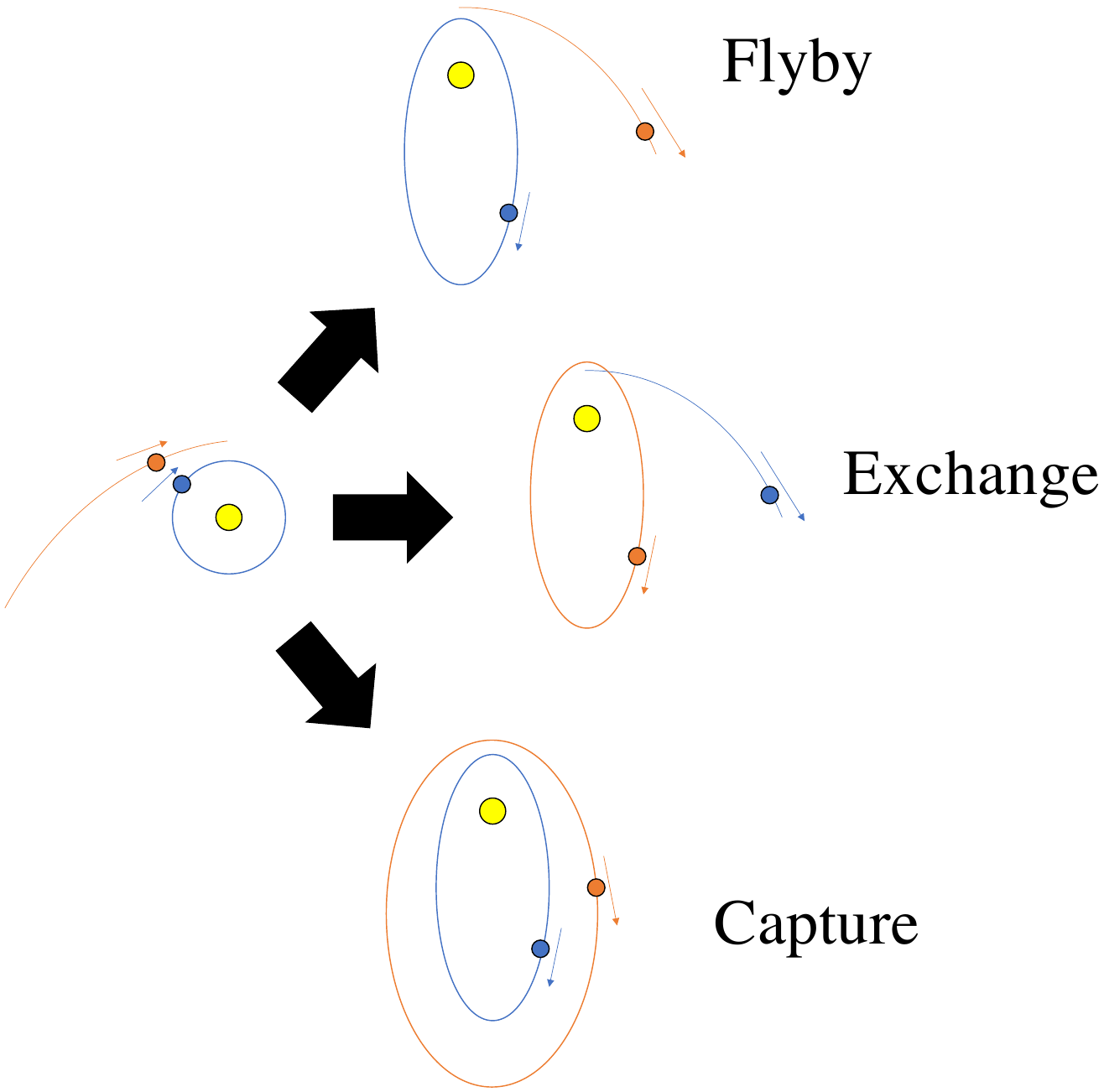}
    \caption{Illustration showing the three possible formation scenarios for HD106906. As the traveling object comes in close proximity with the native companion it can: i) scatter the native companion into a wide and inclined orbit and remain unbound, ii) be captured into a wide and inclined orbit while ejecting the native companion, or iii) scatter the native companion while being captured by the system.}
    \label{fig:cand_outcomes}
\end{figure}

We designed a series of simulations which contain a native companion in an initially circular orbit around a $2.71$ $\mathrm{M_{\odot}}$ central star and a traveler object initially on a nearly parabolic (hyperbolic) orbit that passes by close to the system. The native companion and traveler object are both far enough away from the central mass that the effects of the binary's J2 potential are negligible on the relevant timescales. As the two objects come into close proximity with one another, three outcomes are possible. First, the traveler object can perturb the native companion and excite its eccentricity and inclination and continue on unbound from the system. Alternatively, a close encounter with the traveling object could cause the native companion to become unbound while the traveling object remains captured in an exchange. Finally, the close encounter could lead to both objects being captured onto eccentric and inclined orbits. Figure \ref{fig:cand_outcomes} illustrates the possible outcomes of the free-floating planet flyby.

We selected the semi-major axis of the native companion from a Gaussian distribution centered at $510$ au with a standard deviation of $50$ au, corresponding to the median observed value of the pericenter distance of HD106906b. The masses of each object are also selected from a Gaussian distribution centered at $11$ $\mathrm{M_{Jup}}$ with a standard deviation of $1$ $\mathrm{M_{Jup}}$.

The pericenter distance of the traveling object's initial hyperbolic orbit is also selected from a Gaussian distribution centered at $510$ au with a standard deviation of $50$ au so that the two objects come into relatively close proximity with one another during the flyby. The traveling object's excess velocity (of its initially hyperbolic orbit) is selected from a log-normal distribution on the range $0.1-2$km/s to ensure a nearly parabolic orbit. The inclination of the traveling object's orbit is selected from a sine distribution and the argument of pericenter of the orbit is chosen uniformly randomly so that the encounter geometry is isotropic. Each simulation begins with the traveling object starting at a distance of $10000$ au and ends at the time when the traveling object would return to this distance from the system if it remained unperturbed on its hyperbolic orbit.

The initial phase of the native companion's orbit is carefully selected so that the planets come into relatively close proximity with one another during the flyby. Specifically, we assume that the traveling object will approach the system on an unperturbed hyperbolic orbit. We then select the initial phase of the native companion so that its true anomaly would be equal to the longitude of ascending node of the traveling object at the time that the traveling object would cross the original plane of the native companion's orbit. In reality, the orbits of both the native and traveling objects are perturbed as the traveling object approaches, but this method serves as a good approximation to ensure a close encounter.

For each set of initial conditions that are generated this way, we calculate the distance of closest approach between the two objects under the assumption that their orbits will remain unperturbed. If the distance of closest approach is greater than $50$ au (approximately one Hill radius of the native companion), then the initial conditions are rejected and new ones are generated. This process was repeated until we obtained $100000$ initial conditions.

\begin{figure*}[htb]
\centering
    \includegraphics[width=0.97\textwidth]{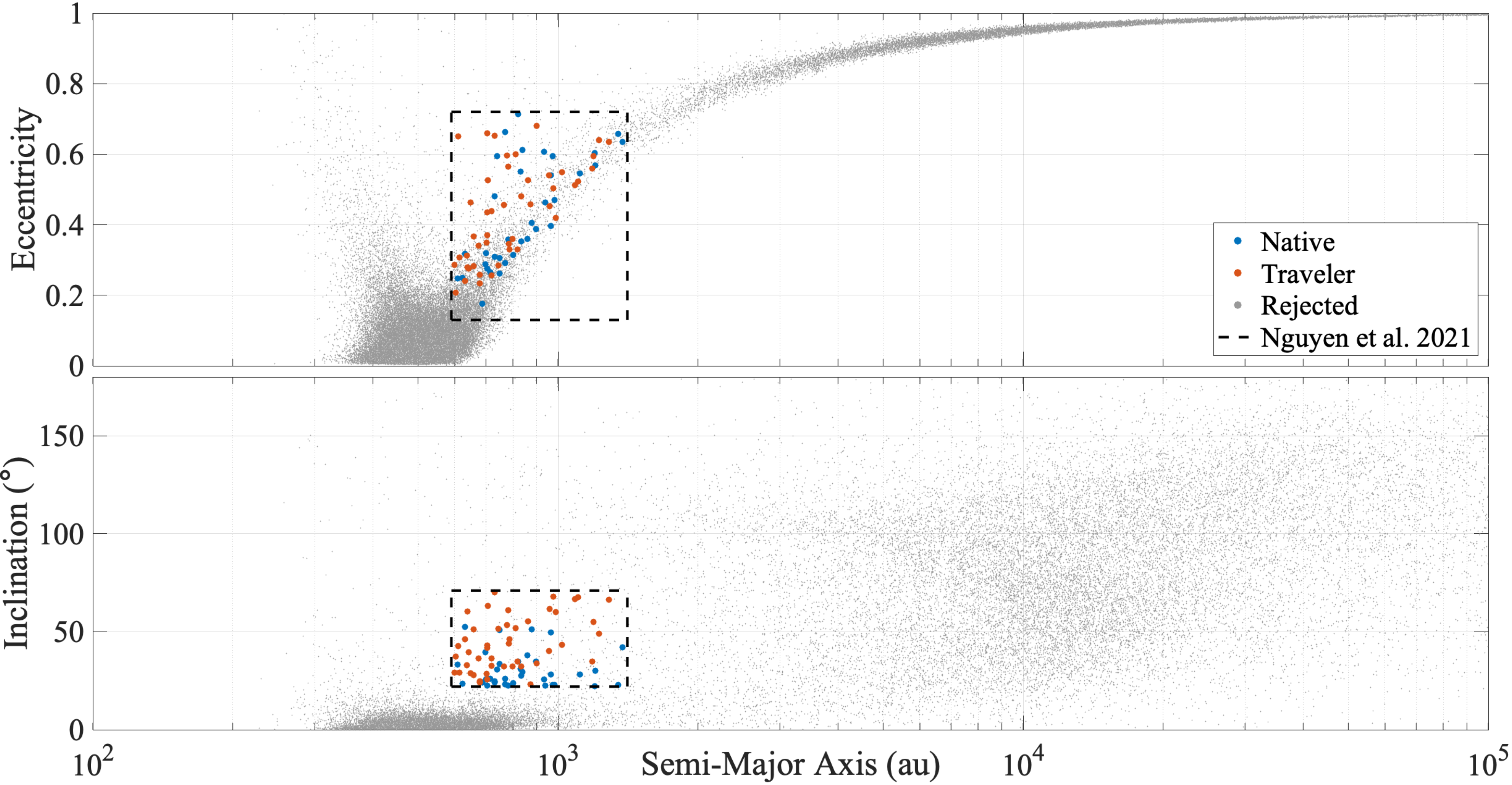}
    \caption{Orbital elements of bound objects after a close encounter between the native companion and the traveler object. Dots in blue denote cases in which the native companion is excited into a favorable configuration that agrees with observation. Dots in orange denote cases in which the traveler object is captured into a favorable configuration. Light grey dots show the distribution of orbital elements from all simulations. The dotted black line shows the observational constrains set by \cite{nguyen_first_2021}.}
    \label{fig:cand_snap_cap}
\end{figure*}

Once the simulations were run to completion, we recorded the orbital elements of the objects that remain bound to the system. These elements are plotted in Figure \ref{fig:cand_snap_cap}.

Figure \ref{fig:cand_snap_cap} shows the semi-major axis, eccentricity, and inclination of objects that remain bound to the system after a close encounter. Dots that are colored show surviving companions that agree with observations of HD106906b. The specific color of the dot denotes whether the planet is native to the system or is a captured traveling object. In the cases in which both objects remain bound to the system, the dots are only colored if one of the objects has a semi-major axis greater than $2000$ au. In these cases we assume that the outer object is only lightly bound to the system and could become unbound in the future with only slight perturbations. The observational constraints from \cite{nguyen_first_2021} are shown with the dashed black line.

As Figure \ref{fig:cand_snap_cap} shows, a close encounter with a free-floating planet is able to produce companions with an eccentricity and or inclination that match observations of HD106906b. The companions could either have their eccentricity and inclination excited by a passing perturber, or a free-floating planet could be captured into an eccentric and inclined orbit that agree with observation.

\subsection{Probability of Close Encounter}
We can make a rough estimate for the rate of encounters between HD106906 and free-floating 11 $\mathrm{M_{Jup}}$ objects, which can be seen as low-mass brown dwarfs. In general, the rate of encounters is given by
\begin{equation}
    \Gamma=n\langle\sigma v\rangle \ ,
    \label{eq:ce_rate}
\end{equation}
where n is the number density of 11 $\mathrm{M_{Jup}}$ objects, $\sigma$ is the encounter cross-section of HD106906, and $v$ is the relative velocity between the system and the 11 $\mathrm{M_{Jup}}$ object.

The number density $n$ can be expressed as a fraction of the stellar number density $n_*$. The stellar number density of LCC, where HD106906 resides, is poorly defined since the density changes as a function of location within the cluster and multiple censuses continually update the boundaries and membership criteria of the cluster. However, since Sco-Cen is the nearest OB association to our Sun, we estimate the average $n_*$ to be slightly more dense than $\sim0.1 \mathrm{pc}^{-3}$, the stellar number density of our solar neighborhood. According to a recent survey of free-floating planets in Upper Scorpius \citep{miret-roig_rich_2021}, we expect there to be 2 to 3 stars per 11 $\mathrm{M_{Jup}}$ object, so we can write
\begin{equation}
    n=A n_* \ ,
    \label{eq:number_dens}
\end{equation}
where $A$ is a dimensionless constant slightly less than unity. Therefore, we roughly estimate $n\sim0.1 \mathrm{pc}^{-3}$.

The cross-section $\sigma$ of HD106906 is enhanced by gravitational focusing and can be expressed as
\begin{equation}
    \sigma=\pi r^2 \left(1+\frac{2 G M}{r v^2}\right) \ ,
    \label{eq:cross_section}
\end{equation}
where $r\sim510$ au is the semi-major axis of the native planet where we expect the close encounter to occur, and $M=2.71$ $\mathrm{M_\odot}$.

We assume the relative velocities follow a Maxwellian distribution $f(v)$ given as
\begin{equation}
    f(v)=\sqrt{\frac{2}{\pi}}\frac{v^2}{s^3}\exp{\left(-\frac{v^2}{2s^2}\right)} \ .
    \label{eq:vel_dis}
\end{equation}
Here we use $s=1.21$ km/s, the 1D velocity dispersion of the Lower Centaurus Crux where HD106906 is located \citep{wright_kinematics_2018}.

The encounter rate can now be expressed as $\Gamma=\Gamma_0 I$
where $\Gamma_0$ is the baseline encounter rate given by
\begin{equation}
    \Gamma_0=A n_* r^2 s \sqrt{2\pi}\approx1.4\times10^{-8}\mathrm{Myr}^{-1} \ ,
    \label{eq:baseline_rate}
\end{equation}
and $I$ is the integral
\begin{equation}
    I = \int_0^\infty\left(1+\frac{2 G M}{r v^2}\right)\frac{v^3}{s^3}\exp{\left(-\frac{v^2}{2s^2}\right)}\frac{dv}{s}\approx6400 \ .
    \label{eq:I_int}
\end{equation}

Since we require that the incoming object passes a bit inside the Hill Sphere of the native companion, we multiply the encounter rate by an overall factor $g=B [2r_H/(2\pi r)][2r_H/r]$, where $r_H$ is the radius of the native companion's Hill Sphere and $B$ is a dimensionless constant of order unity. The encounter rate can now be evaluated as
\begin{equation}
    \Gamma = g\Gamma_0 I \sim 10^{-4} \mathrm{Myr}^{-1} \ .
    \label{eq:ce_rate_final}
\end{equation}

Based upon our simulations in Section \ref{sec:scattering}, we expect that a close encounter of this type will produce a companion within observational constraints $\sim0.2\%$ of the time. We also showed in Section \ref{sec:vert_warp} and Section \ref{sec:smack} that once a companion establishes an eccentric and inclined orbit around the disk, the disk will become severely vertically warped after only $\sim5$ Myr. Therefore, we estimate the probability that this scenario occurred to be
\begin{equation}
    P=0.002\times\Gamma\times5\mathrm{Myr}\sim10^{-6} \ .
    \label{eq:prob}
\end{equation}
This formation probability is quite low, which is expected for a system which has such an unusual configuration. If this probability were higher we would expect to observe many more systems that are similar to HD106906, but its uniqueness causes it to stand out. Despite the low value, our probability is actually high compared to previous formation theories for this system, roughly an order of magnitude greater \citep{rodet_origin_2017}.

The least constrained value in this calculation is $n$, the number density of $11$ $\mathrm{M_{Jup}}$ objects. According to the results from \cite{miret-roig_rich_2021}, the estimation of the relative number of these objects can vary by as much as $50\%$. Since the encounter rate that we calculate is directly proportional to $n$, this variation introduces significant uncertainty into our calculation. However, at its lowest value, $P\sim5\times10^{-7}$, the probability of this formation theory is about twice as likely as those previously proposed.

The calculation is more complicated if we consider a close encounter between HD106906 and objects with either a higher or lower mass than $11$ $\mathrm{M_{Jup}}$. If the traveling object has lower mass, then the probability of this scenario decreases. According to \cite{miret-roig_rich_2021}, we expect that the density of free-floating lower mass objects (that are still large enough to significantly alter the orbit of a native companion to HD106906) to be lower than $11$ $\mathrm{M_{Jup}}$ objects. In addition, if a traveler has lower mass, it is less likely to scatter a native object into a configuration that agrees with observations of HD106906b. On the other hand, more massive objects, like brown dwarfs, are typically more common than free-floating planets. HD106906 likely has a higher encounter rate with these more massive objects. More massive perturbers may also more easily scatter a native object into a favorable configuration. However, if the traveling object has a mass that is much different than HD106906b, then we must discount all of the `Exchange' and some of the `Capture' outcomes of our scenario since we have a good estimate of the mass of the surviving object. Discounting these outcomes decreases the overall probability.

We note that the companion HD106906b is bright enough to be detected due to its young age. A captured free floating object may miss detection if it is much older and cooler. However, if the free floating object originated in the cluster then it is also likely young since the mean age of the cluster is only $\sim17$ Myr. Specifically, objects that are $\sim13$ $\mathrm{M_{Jup}}$ and $17$ Myr old are similar in brightness to HD106906b (assuming it is native with $11$ $\mathrm{M_{Jup}}$ and $13$ Myr old) \citep{chabrier_evolutionary_2000,baraffe_evolutionary_2003,bailey_hd_2014}. Therefore we do not expect the detectability of free floating objects in the cluster to affect our results significantly.

\section{Conclusion}\label{sec:conc}
The unusual structure of the system HD106906 presents challenges to our classical understanding of how planetary systems form. The massive companion at such a large separation from its host star is itself unusual and the eccentric and inclined nature of its orbit requires an explanation. The relatively flat debris disk interior to the companion's orbit allows us to place constraints upon the formation history of the system as any formation scenario would have to recreate and preserve this structure.

It has been shown previously that the eccentricity of the outer disk can be excited through external perturbations by the companion \citep{jilkova_debris_2015,nesvold_hd_2017}. However, the relatively high inclination of the companion ($>35^\circ$), recently constrained by observation \citep{nguyen_first_2021}, is likely to warp the vertical structure of the disk on a relatively short timescale. Our results in Section \ref{sec:vert_warp} show that, in the presence of the eccentric and inclined companion, the vertical structure of the disk will warp such that the vertical FWHM of the debris disk particles will exceed observational constraints \citep{crotts_deep_2021} by $\sim1-5$ Myr or sooner, depending upon the exact orbital parameters of the companion. The eccentricity and inclination excitation of the debris disk particles is driven by close encounter scattering in some cases and secular perturbations from the companion in most cases. Figure \ref{fig:disk_evo_params} shows that the vertical warping timescale of the disk in the presence of a companion with orbital parameters that agree with observational constraint is on the order of $\lesssim5$ Myr.

In Section \ref{sec:smack}, we studied in closer detail a model system with a companion that has a semi-major axis $850$ au, eccentricity $0.4$, and mutual inclination with the disk of $40^\circ$. We then produced simulated observational images of the inner disk that results from this configuration. These simulated observational images were constructed by modeling dust producing collisions between planetesimals within the disk, and then evolving the orbits of the dust particles under the influence of gravitational forces and radiation pressure from the central binary. The surface brightness of the dust is then calculated based upon its location and $\beta$ value (which is related to the size of each dust grain). The resulting simulated observational images are shown in comparison to real observed images of HD106906 in Figure \ref{fig:smack_images}. This eccentric and inclined configuration of the companion allows the outer disk eccentricity to grow while preserving the relative flatness of the inner disk. This configuration also produces a simulated image that is agreeable with real observed images of the inner disk of the system at $1$ Myr of evolution. We also showed that, by $5$ Myr, the puffiness of the disk exceeds observational constraints on its flatness.

To date, HD106906 remains the only known system containing a debris disk with a directly imaged external companion. Other debris disks whose vertical structure has become warped or puffy might indicate the presence of an undiscovered inclined external companion that is torquing the disk on long timescales. We speculate that similar features in the debris disks HIP 79977 and AU Microscopii may also indicate a possible external companion \citep{boccaletti_fast-moving_2015,engler_hip_2017}.

Our findings suggest that the inclination excitation of the companion to HD106906 must have happened relatively recently, since the flat structure of the disk cannot be preserved for long. This recent timing requirement places significant constraints upon any proposed formation scenario for the system. We thus proposed a formation scenario that the system formed with a planetary mass companion in the plane of its debris disk. The companion could have formed in the disk around the proto-binary, or from a portion of a self-gravitating molecular cloud which fragmented into a binary and companion. Then, a recent close encounter with a free-floating planet could have scattered the native companion. If HD106906b is indeed a result of a flyby that is non-native to the system, this could help explain the misalignment of its spin axis with its orbit normal \citep{bryan_obliquity_2021}. The eccentric and inclined orbit of the surviving companion is able to excite the eccentricity of the outer disk giving rise to the observed asymmetry.

A recent survey of free-floating planets in USCO suggests that the relative abundance of these objects has been previously underestimated \citep{miret-roig_rich_2021}. This opens up the possibility that these objects may be significant in the early dynamics of young solar systems which still reside within their birth cluster. Their typically low brightness would also mean they could escape detection by observational surveys. Their low mass would also mean that close encounters between these objects and native planets could significantly perturb planetary systems while leaving the structure of debris disks relatively unchanged. However, the low mass of these objects also means that for a close encounter to significantly alter the orbit of native planets, the distance of closest approach between the free-floating planet and native companion must be quite small ($\sim1r_H$).

In Section \ref{sec:scattering}, we modeled this close encounter scenario with a free-floating planet using $100000$ initial conditions. We showed that this scenario is able to reproduce a surviving companion with the orbital elements that fall within observational constraints. We also showed that it is possible to produce a highly eccentric and highly inclined companion that we suggest is necessary to grow the eccentricity of the disk while also minimizing its vertical warp. In the end, we estimated the probability that a close encounter of this type occurred within the last $5$ Myr, and we found a probability of $~10^{-6}$, which is roughly an order of magnitude higher than probabilities of previous formation theories for this system.

\begin{acknowledgments}
We thank Dr. Meredith Hughes for helpful discussions on this system. GL and NM are grateful for the support by NASA 80NSSC20K0641 and 80NSSC20K0522. This work used the Hive cluster, which is supported by the National Science Foundation under grant number 1828187. SN acknowledges the partial support from NASA ATP AWD-000836-G1 and thanks Howard and Astrid Preston for their generous support. This research was supported in part through research cyberinfrastrucutre resources and services provided by the Partnership for an Advanced Computing Environment (PACE) at the Georgia Institute of Technology, Atlanta, Georgia, USA.
\end{acknowledgments}

\bibliographystyle{hapj.bst}
\bibliography{msref.bib}

\begin{thebibliography}{54}
\expandafter\ifx\csname natexlab\endcsname\relax\def\natexlab#1{#1}\fi

\bibitem[{Ardila {et~al.}(2005)Ardila, Lubow, Golimowski, Krist, Clampin, Ford,
  Hartig, Illingworth, Bartko, BenÍtez, Blakeslee, Bouwens, Bradley,
  Broadhurst, Brown, Burrows, Cheng, Cross, Feldman, Franx, Goto, Gronwall,
  Holden, Homeier, Infante, Kimble, Lesser, Martel, Menanteau, Meurer, Miley,
  Postman, Sirianni, Sparks, Tran, Tsvetanov, White, Zheng, \&
  Zirm}]{ardila_dynamical_2005}
Ardila, D.~R. {et~al.} 2005, The Astrophysical Journal, 627, 986

\bibitem[{Bailey {et~al.}(2014)Bailey, Meshkat, Reiter, Morzinski, Males, Su,
  Hinz, Kenworthy, Stark, Mamajek, Briguglio, Close, Follette, Puglisi,
  Rodigas, Weinberger, \& Xompero}]{bailey_hd_2014}
Bailey, V. {et~al.} 2014, The Astrophysical Journal Letters, 780, L4

\bibitem[{Ballering {et~al.}(2016)Ballering, Su, Rieke, \&
  Gáspár}]{ballering_comprehensive_2016}
Ballering, N.~P., Su, K. Y.~L., Rieke, G.~H., \& Gáspár, A. 2016, The
  Astrophysical Journal, 823, 108

\bibitem[{Baraffe {et~al.}(2003)Baraffe, Chabrier, Barman, Allard, \&
  Hauschildt}]{baraffe_evolutionary_2003}
Baraffe, I., Chabrier, G., Barman, T.~S., Allard, F., \& Hauschildt, P.~H.
  2003, Astronomy and Astrophysics, 402, 701

\bibitem[{Bate(2009)}]{bate_stellar_2009}
Bate, M.~R. 2009, Monthly Notices of the Royal Astronomical Society, 392, 590

\bibitem[{Bate(2018)}]{bate_diversity_2018}
------. 2018, Monthly Notices of the Royal Astronomical Society, 475, 5618

\bibitem[{Bate {et~al.}(2002)Bate, Bonnell, \& Bromm}]{bate_formation_2002}
Bate, M.~R., Bonnell, I.~A., \& Bromm, V. 2002, Monthly Notices of the Royal
  Astronomical Society, 332, L65

\bibitem[{Batygin {et~al.}(2020)Batygin, Adams, Batygin, \&
  Petigura}]{batygin_dynamics_2020}
Batygin, K., Adams, F.~C., Batygin, Y.~K., \& Petigura, E.~A. 2020, The
  Astronomical Journal, 159, 101

\bibitem[{Boccaletti {et~al.}(2015)Boccaletti, Thalmann, Lagrange, Janson,
  Augereau, Schneider, Milli, Grady, Debes, Langlois, Mouillet, Henning,
  Dominik, Maire, Beuzit, Carson, Dohlen, Engler, Feldt, Fusco, Ginski, Girard,
  Hines, Kasper, Mawet, Ménard, Meyer, Moutou, Olofsson, Rodigas, Sauvage,
  Schlieder, Schmid, Turatto, Udry, Vakili, Vigan, Wahhaj, \&
  Wisniewski}]{boccaletti_fast-moving_2015}
Boccaletti, A. {et~al.} 2015, Nature, 526, 230

\bibitem[{Boss(1997)}]{boss_giant_1997}
Boss, A.~P. 1997, Science, 276, 1836

\bibitem[{Bryan {et~al.}(2020)Bryan, Chiang, Bowler, Morley, Millholland,
  Blunt, Ashok, Nielsen, Ngo, Mawet, \& Knutson}]{bryan_obliquity_2020}
Bryan, M.~L. {et~al.} 2020, The Astronomical Journal, 159, 181

\bibitem[{Bryan {et~al.}(2021)Bryan, Chiang, Morley, Mace, \&
  Bowler}]{bryan_obliquity_2021}
Bryan, M.~L., Chiang, E., Morley, C.~V., Mace, G.~N., \& Bowler, B.~P. 2021,
  The Astronomical Journal, 162, 217

\bibitem[{Cai {et~al.}(2019)Cai, Portegies~Zwart, Kouwenhoven, \&
  Spurzem}]{cai_survivability_2019}
Cai, M.~X., Portegies~Zwart, S., Kouwenhoven, M. B.~N., \& Spurzem, R. 2019,
  Monthly Notices of the Royal Astronomical Society, 489, 4311

\bibitem[{Chabrier {et~al.}(2000)Chabrier, Baraffe, Allard, \&
  Hauschildt}]{chabrier_evolutionary_2000}
Chabrier, G., Baraffe, I., Allard, F., \& Hauschildt, P. 2000, The
  Astrophysical Journal, 542, 464

\bibitem[{Chambers(2012)}]{chambers_mercury_2012}
Chambers, J.~E. 2012, Astrophysics Source Code Library, ascl:1201.008

\bibitem[{Crotts {et~al.}(2021)Crotts, Matthews, Esposito, Duchêne, Kalas,
  Chen, Arriaga, Millar-Blanchaer, Debes, Draper, Fitzgerald, Hom, MacGregor,
  Mazoyer, Patience, Rice, Weinberger, Wilner, \& Wolff}]{crotts_deep_2021}
Crotts, K.~A. {et~al.} 2021, The Astrophysical Journal, 915, 58

\bibitem[{De~Rosa \& Kalas(2019)}]{de_rosa_near-coplanar_2019}
De~Rosa, R.~J., \& Kalas, P. 2019, The Astronomical Journal, 157, 125

\bibitem[{de~Zeeuw {et~al.}(1999)de~Zeeuw, Hoogerwerf, de~Bruijne, Brown, \&
  Blaauw}]{de_zeeuw_hipparcos_1999}
de~Zeeuw, P.~T., Hoogerwerf, R., de~Bruijne, J. H.~J., Brown, A. G.~A., \&
  Blaauw, A. 1999, The Astronomical Journal, 117, 354

\bibitem[{Engler {et~al.}(2017)Engler, Schmid, Thalmann, Boccaletti, Bazzon,
  Baruffolo, Beuzit, Claudi, Costille, Desidera, Dohlen, Dominik, Feldt, Fusco,
  Ginski, Gisler, Girard, Gratton, Henning, Hubin, Janson, Kasper, Kral,
  Langlois, Lagadec, Ménard, Meyer, Milli, Mouillet, Olofsson, Pavlov, Pragt,
  Puget, Quanz, Roelfsema, Salasnich, Siebenmorgen, Sissa, Suarez, Szulagyi,
  Turatto, Udry, \& Wildi}]{engler_hip_2017}
Engler, N. {et~al.} 2017, Astronomy \& Astrophysics, 607, A90

\bibitem[{Farhat {et~al.}(2022)Farhat, Sefilian, \& Touma}]{farhat_case_2022}
Farhat, M., Sefilian, A., \& Touma, J. 2022, The case of {HD} 106906 debris
  disc: {A} binary's revenge, arXiv:2210.07395 [astro-ph]

\bibitem[{Fehr {et~al.}(2022)Fehr, Hughes, Dawson, Marino, Ackelsberg,
  Kittling, Flaherty, Nesvold, Carpenter, Andrews, Matthews, Crotts, \&
  Kalas}]{fehr_millimeter_2022}
Fehr, A.~J. {et~al.} 2022, The Astrophysical Journal, 939, 56

\bibitem[{Henyey \& Greenstein(1941)}]{henyey_diffuse_1941}
Henyey, L.~G., \& Greenstein, J.~L. 1941, The Astrophysical Journal, 93, 70

\bibitem[{Hughes {et~al.}(2018)Hughes, Duchêne, \&
  Matthews}]{hughes_debris_2018}
Hughes, A.~M., Duchêne, G., \& Matthews, B.~C. 2018, Annual Review of
  Astronomy and Astrophysics, 56, 541

\bibitem[{Jennings \& Chiang(2021)}]{jennings_primordial_2021}
Jennings, R.~M., \& Chiang, E. 2021, Monthly Notices of the Royal Astronomical
  Society, 507, 5187

\bibitem[{Jílková \& Portegies~Zwart(2015)}]{jilkova_debris_2015}
Jílková, L., \& Portegies~Zwart, S. 2015, Monthly Notices of the Royal
  Astronomical Society, 451, 804

\bibitem[{Kalas {et~al.}(2007)Kalas, Fitzgerald, \&
  Graham}]{kalas_discovery_2007}
Kalas, P., Fitzgerald, M.~P., \& Graham, J.~R. 2007, The Astrophysical Journal
  Letters, 661, L85

\bibitem[{Kalas \& Jewitt(1995)}]{kalas_asymmetries_1995}
Kalas, P., \& Jewitt, D. 1995, The Astronomical Journal, 110, 794

\bibitem[{Kalas {et~al.}(2015)Kalas, Rajan, Wang, Millar-Blanchaer, Duchene,
  Chen, Fitzgerald, Dong, Graham, Patience, Macintosh, Murray-Clay, Matthews,
  Rameau, Marois, Chilcote, De~Rosa, Doyon, Draper, Lawler, Ammons, Arriaga,
  Bulger, Cotten, Follette, Goodsell, Greenbaum, Hibon, Hinkley, Hung,
  Ingraham, Konapacky, Lafreniere, Larkin, Long, Maire, Marchis, Metchev,
  Morzinski, Nielsen, Oppenheimer, Perrin, Pueyo, Rantakyrö, Ruffio,
  Saddlemyer, Savransky, Schneider, Sivaramakrishnan, Soummer, Song, Thomas,
  Vasisht, Ward-Duong, Wiktorowicz, \& Wolff}]{kalas_direct_2015}
Kalas, P.~G. {et~al.} 2015, The Astrophysical Journal, 814, 32

\bibitem[{Kinoshita \& Nakai(1999)}]{kinoshita_analytical_1999}
Kinoshita, H., \& Nakai, H. 1999, Celestial Mechanics and Dynamical Astronomy,
  75, 125

\bibitem[{Lagrange {et~al.}(2016)Lagrange, Langlois, Gratton, Maire, Milli,
  Olofsson, Vigan, Bailey, Mesa, Chauvin, Boccaletti, Galicher, Girard,
  Bonnefoy, Samland, Menard, Henning, Kenworthy, Thalmann, Beust, Beuzit,
  Brandner, Buenzli, Cheetham, Janson, le~Coroller, Lannier, Mouillet, Peretti,
  Perrot, Salter, Sissa, Wahhaj, Abe, Desidera, Feldt, Madec, Perret, Petit,
  Rabou, Soenke, \& Weber}]{lagrange_narrow_2016}
Lagrange, A.-M. {et~al.} 2016, Astronomy and Astrophysics, 586, L8

\bibitem[{Lee \& Chiang(2016)}]{lee_primer_2016}
Lee, E.~J., \& Chiang, E. 2016, The Astrophysical Journal, 827, 125

\bibitem[{Li {et~al.}(2019)Li, Mustill, \& Davies}]{li_fly-by_2019}
Li, D., Mustill, A.~J., \& Davies, M.~B. 2019, Monthly Notices of the Royal
  Astronomical Society, 488, 1366

\bibitem[{Li {et~al.}(2020)Li, Mustill, \& Davies}]{li_flyby_2020}
------. 2020, Monthly Notices of the Royal Astronomical Society, 496, 1149

\bibitem[{Lieman-Sifry {et~al.}(2016)Lieman-Sifry, Hughes, Carpenter, Gorti,
  Hales, \& Flaherty}]{lieman-sifry_debris_2016}
Lieman-Sifry, J., Hughes, A.~M., Carpenter, J.~M., Gorti, U., Hales, A., \&
  Flaherty, K.~M. 2016, The Astrophysical Journal, 828, 25

\bibitem[{Matthews {et~al.}(2014)Matthews, Krivov, Wyatt, Bryden, \&
  Eiroa}]{matthews_observations_2014}
Matthews, B.~C., Krivov, A.~V., Wyatt, M.~C., Bryden, G., \& Eiroa, C. 2014,
  Observations, {Modeling}, and {Theory} of {Debris} {Disks} (eprint:
  arXiv:1401.0743), conference Name: Protostars and Planets VI Pages: 521 ADS
  Bibcode: 2014prpl.conf..521M

\bibitem[{Maury {et~al.}(2019)Maury, André, Testi, Maret, Belloche,
  Hennebelle, Cabrit, Codella, Gueth, Podio, Anderl, Bacmann, Bontemps, Gaudel,
  Ladjelate, Lefèvre, Tabone, \& Lefloch}]{maury_characterizing_2019}
Maury, A.~J. {et~al.} 2019, Astronomy and Astrophysics, 621, A76

\bibitem[{Miret-Roig {et~al.}(2021)Miret-Roig, Bouy, Raymond, Tamura, Bertin,
  Barrado, Olivares, Galli, Cuillandre, Sarro, Berihuete, \&
  Huélamo}]{miret-roig_rich_2021}
Miret-Roig, N. {et~al.} 2021, Nature Astronomy

\bibitem[{Naoz(2016)}]{naoz_eccentric_2016}
Naoz, S. 2016, Annual Review of Astronomy and Astrophysics, 54, 441

\bibitem[{Naoz {et~al.}(2013)Naoz, Farr, Lithwick, Rasio, \&
  Teyssandier}]{naoz_secular_2013}
Naoz, S., Farr, W.~M., Lithwick, Y., Rasio, F.~A., \& Teyssandier, J. 2013,
  Monthly Notices of the Royal Astronomical Society, 431, 2155

\bibitem[{Nesvold {et~al.}(2013)Nesvold, Kuchner, Rein, \&
  Pan}]{nesvold_smack_2013}
Nesvold, E.~R., Kuchner, M.~J., Rein, H., \& Pan, M. 2013, The Astrophysical
  Journal, 777, 144

\bibitem[{Nesvold {et~al.}(2017)Nesvold, Naoz, \& Fitzgerald}]{nesvold_hd_2017}
Nesvold, E.~R., Naoz, S., \& Fitzgerald, M.~P. 2017, The Astrophysical Journal
  Letters, 837, L6

\bibitem[{Nesvold {et~al.}(2016)Nesvold, Naoz, Vican, \&
  Farr}]{nesvold_circumstellar_2016}
Nesvold, E.~R., Naoz, S., Vican, L., \& Farr, W.~M. 2016, The Astrophysical
  Journal, 826, 19

\bibitem[{Nguyen {et~al.}(2021)Nguyen, De~Rosa, \& Kalas}]{nguyen_first_2021}
Nguyen, M.~M., De~Rosa, R.~J., \& Kalas, P. 2021, The Astronomical Journal,
  161, 22

\bibitem[{Pollack {et~al.}(1996)Pollack, Hubickyj, Bodenheimer, Lissauer,
  Podolak, \& Greenzweig}]{pollack_formation_1996}
Pollack, J.~B., Hubickyj, O., Bodenheimer, P., Lissauer, J.~J., Podolak, M., \&
  Greenzweig, Y. 1996, Icarus, 124, 62

\bibitem[{Reche {et~al.}(2008)Reche, Beust, Augereau, \&
  Absil}]{reche_observability_2008}
Reche, R., Beust, H., Augereau, J.-C., \& Absil, O. 2008, Astronomy and
  Astrophysics, 480, 551

\bibitem[{Rein \& Liu(2012)}]{rein_rebound_2012}
Rein, H., \& Liu, S.-F. 2012, Astronomy and Astrophysics, 537, A128

\bibitem[{Rodet {et~al.}(2019)Rodet, Beust, Bonnefoy, De~Rosa, Kalas, \&
  Lagrange}]{rodet_odea_2019}
Rodet, L., Beust, H., Bonnefoy, M., De~Rosa, R.~J., Kalas, P., \& Lagrange,
  A.-M. 2019, Astronomy and Astrophysics, 631, A139

\bibitem[{Rodet {et~al.}(2017)Rodet, Beust, Bonnefoy, Lagrange, Galli,
  Ducourant, \& Teixeira}]{rodet_origin_2017}
Rodet, L., Beust, H., Bonnefoy, M., Lagrange, A.-M., Galli, P. A.~B.,
  Ducourant, C., \& Teixeira, R. 2017, Astronomy and Astrophysics, 602, A12

\bibitem[{Wagner {et~al.}(2018)Wagner, Dong, Sheehan, Apai, Kasper, McClure,
  Morzinski, Close, Males, Hinz, Quanz, \& Fung}]{wagner_orbit_2018}
Wagner, K. {et~al.} 2018, The Astrophysical Journal, 854, 130

\bibitem[{Wang {et~al.}(2020)Wang, Perna, \& Leigh}]{wang_planetary_2020}
Wang, Y.-H., Perna, R., \& Leigh, N. W.~C. 2020, Monthly Notices of the Royal
  Astronomical Society, 496, 1453

\bibitem[{Wright \& Mamajek(2018)}]{wright_kinematics_2018}
Wright, N.~J., \& Mamajek, E.~E. 2018, Monthly Notices of the Royal
  Astronomical Society, 476, 381

\bibitem[{Wu {et~al.}(2016)Wu, Close, Bailey, Rodigas, Males, Morzinski,
  Follette, Hinz, Puglisi, Briguglio, \& Xompero}]{wu_magellan_2016}
Wu, Y.-L. {et~al.} 2016, The Astrophysical Journal, 823, 24

\bibitem[{Wyatt(2008)}]{wyatt_evolution_2008}
Wyatt, M.~C. 2008, Annual Review of Astronomy and Astrophysics, 46, 339

\bibitem[{Wyatt {et~al.}(1999)Wyatt, Dermott, Telesco, Fisher, Grogan, Holmes,
  \& Piña}]{wyatt_how_1999}
Wyatt, M.~C., Dermott, S.~F., Telesco, C.~M., Fisher, R.~S., Grogan, K.,
  Holmes, E.~K., \& Piña, R.~K. 1999, The Astrophysical Journal, 527, 918

\end{thebibliography}

\end{document}